\documentstyle[prl,multicol,aps]{revtex}
\begin{document}
\draft
\title{Decay properties of the density matrix and Wannier functions for
       interacting systems}
\author{Stefan Goedecker and Erik Koch}
\address{Max-Planck Institut f\"ur Festk\"orperforschung, 
         Heisenbergstra\ss e 1, 70569 Stuttgart, Germany}
\date{\today}
\maketitle
\begin{abstract}
For non-interacting electrons the one-particle density matrix and the related
Wannier functions characterize a material as insulating or metallic. 
Introducing many-body Wannier functions, we show that this characterization 
can be carried over to interacting systems. In particular, we find that an 
exponential decay of the density matrix characterizes not only band insulators 
but also Mott insulators. The properties of the many-body Wannier functions 
differ, however, from those of the Wannier functions of a non-interacting 
systems.
\end{abstract}

\begin{multicols}{2}
\setcounter{collectmore}{5}
\raggedcolumns
The locality properties of solids within independent-particle theories 
have been at the focus of much attention \cite{onepart}. It was shown that 
for insulators the density matrix decays exponentially. The rate of the decay 
is related to the size of the energy gap in the band-structure: A larger gaps 
mean a faster decay. In contrast, for a metal at zero temperature, the decay 
is much slower, namely algebraic. 
The decay properties of the density matrix are reflected in the locality
properties of the Wannier functions \cite{ashcroft,blount}. The Wannier
function $w_j({\bf r}-{\bf R})$ associated with the $j$-th band is localized 
around the lattice vector ${\bf R}$, i.e.\ it tends to zero for ${\bf r}$
far from ${\bf R}$. In an insulator this decay is exponential for isolated 
bands \cite{kohnwan}.
The practical importance of the Wannier functions is two-fold. First, they
allow for an intuitive interpretation of polarization effects and bonding 
properties in a solid \cite{marzari}: Localized Wannier functions correspond 
to either bonds or lone electron pairs. Second, they are the basic quantity 
in some linear-scaling algorithms \cite{myrev,ordrev}. When they are localized,
they need only be calculated within the localization region, allowing thus to
use divide and conquer strategies that finally lead to linear scaling.

In the following we will show how the characterization of an insulator by
the exponential decay of the density matrix can be carried over to interacting
systems. Using natural orbitals and what 
we call many-body Wannier functions allows us to discuss these many-body 
properties in terms of single particle wave-functions. It turns out 
that the localization properties of the occupied many-body Wannier 
functions are similar to those of the Wannier functions for non-interacting 
systems. For the virtual Wannier functions there are, however, qualitative 
differences.

Criteria for characterizing the insulating state that are valid beyond the
independent-particle picture have already been put forward. Kohn \cite{kohn}
has shown that the many-body wavefunction of an insulating ring breaks up
into a sum of functions which are localized in disconnected regions of the
many-particle configuration space. Resta and Sorella \cite{resta} have shown
that the expectation value of the operator $e^{i(2\pi/L)\hat{X}}$ can be used
to distinguish between insulators and metals. This criterion seems, however,
to be restricted to one-dimensional systems. 
As we will show, a more general criterion is given by the decay of the density
matrix. 
All these criteria have in common that they refer only to the ground state
wavefunction and do not require information about excited states. Criteria
of this type are the most useful ones in numerical calculations, where in
general only the ground state is calculated. 

We start our construction by generalizing the concept of Wannier functions to
interacting systems. Let us consider a $N$-electron wavefunction 
$\Psi  ({\bf x}_1, {\bf x}_2, ... ,{\bf x}_N)$, where ${\bf x}= \{{\bf r}, s\}$
is a combined spatial and spin coordinate. The reduced density matrix 
$D({\bf r},{\bf r}')$ is then given by
\begin{eqnarray}  \label{red1}
D({\bf r},{\bf r}') &=& N \int\,ds d{\bf x}_2 ...d{\bf x}_N \nonumber\\
& & \Psi^\ast ({\bf r}, s, {\bf x}_2, ...{\bf x}_N) \:
    \Psi      ({\bf r}',s, {\bf x}_2, ...{\bf x}_N) \; . 
\end{eqnarray}
The eigenfunctions $\Phi_i({\bf r})$ of the reduced density matrix are
called the natural orbitals, its eigenvalues $n_i$ the natural occupation
numbers. In this basis the many-body density matrix takes the same form as 
an independent-particle density matrix:
\begin{displaymath}
D({\bf r},{\bf r}')=\sum_i n_i\:\Phi^\ast_i({\bf r})\:\Phi_i({\bf r}')\:.
\end{displaymath}

Translating all the spatial arguments of $\Psi$ by a lattice vector ${\bf R}$ 
multiplies the wavefunction by a phase factor. Since these phase factors cancel 
by the definition of the density matrix $D$ we find
\begin{displaymath}
D({\bf r}+{\bf R},{\bf r}'+{\bf R}) = D({\bf r},{\bf r}') \: .
\end{displaymath}
$D({\bf r}',{\bf r})$ has thus the same symmetry properties as the Hamiltonian 
of a periodic solid in an independent-particle picture \cite{ashcroft}. 
Therefore its eigenfunctions can be written as Bloch functions, labeled by a 
band index $j$ and a vector ${\bf k}$ contained in the Brillouin zone
\begin{equation} \label{natk} 
\Phi_{j,{\bf k}}({\bf r}) = e^{i{\bf k} {\bf r}} \: U_{j,{\bf k}}({\bf r}) \:.
\end{equation}
The $U_{j,{\bf k}}({\bf r})$ are periodic functions with respect to the real 
space primitive cell. From these functions we construct the ${\bf k}$-dependent
density matrix, which plays a central role in our argument: 
\begin{displaymath}
D_{\bf k}({\bf r},{\bf r}') = \sum_j n_j({\bf k})\: 
                  U_{j,{\bf k}}({\bf r}) \: U^\ast_{j,{\bf k}}({\bf r'}) \:,
\end{displaymath}
where the $n_j({\bf k})$ are the occupation numbers corresponding to the 
natural orbitals (\ref{natk}). 
The full density matrix is related to the ${\bf k}$-dependent density matrix by
\begin{equation} \label{fdm} 
D({\bf r},{\bf r}') = \frac{V}{(2 \pi)^3} \int_{BZ} d{\bf k}\;
    D_{\bf k}({\bf r},{\bf r}') \: e^{i{\bf k} ({\bf r}-{\bf r}')} \:,
\end{equation}
where $V$ is the volume of the real-space primitive cell and the integration
is over the Brillouin Zone (BZ). 
By construction the $U_{j,{\bf k}}({\bf r})$ satisfy the eigenvalue equation
\begin{equation}  \label{eveq}
\int d{\bf r}'\;D_{\bf k}({\bf r},{\bf r}') \: U_{j,{\bf k}}({\bf r}') = 
             n_j({\bf k}) \: U_{j,{\bf k}}({\bf r}) \: .
\end{equation}
Because of the analogy to the band structure problem, we call the eigenvalues
$n_j({\bf k})$ the occupation band-structure. It has the periodicity of the
Brillouin zone, and N-representability \cite{coleman} requires that 
$n_j({\bf k}) \in [0,2]$ ($[0,1]$ without spin degeneracy).

We next use the natural orbitals (\ref{natk}) to define many-body Wannier 
functions, $W_j({\bf r}-{\bf R})$ 
\begin{equation}\label{mbw} 
W_j({\bf r}-{\bf R}) = {V\over(2 \pi)^3} \int_{BZ} d{\bf k}\;
            e^{i {\bf k}({\bf r}-{\bf R})} \: U_{j,{\bf k}}({\bf r}) \:.
\end{equation}
This definition is formally similar to the usual definition of the 
independent-particle Wannier functions:
\begin{displaymath}
w_j({\bf r}-{\bf R}) = {V\over(2 \pi)^3} \int_{BZ} d{\bf k}\;
            e^{i {\bf k}({\bf r}-{\bf R})} \: u_{j,{\bf k}}({\bf r}) \:,
\end{displaymath}
where $\phi_{j,{\bf K}}({\bf r})$ are the independent-particle Bloch functions
\begin{displaymath}
\phi_{j,{\bf K}}({\bf r})=e^{i{\bf k}{\bf r}}\,u_{j,{\bf k}}({\bf r}),
\end{displaymath}
whose cell-periodic part satisfies the eigenvalue problem 
\begin{equation} \label{bnd}
\int d{\bf r}'\; H^0_{\bf k}({\bf r},{\bf r}')\:u_{j,{\bf K}}({\bf r}')
= \epsilon_j({\bf k})\:u_{j,{\bf k}}({\bf r})\:.
\end{equation}
We notice that in the non-interacting case the eigenvalue problem (\ref{eveq})
involving the density matrix is replaced by (\ref{bnd}), involving the 
effective Hamiltonian $H^0_{\bf k}({\bf r},{\bf r}')$. The reason for this
is that in the limit of no interaction the density matrix becomes highly
rank deficient, with zero occupation number for all virtual orbitals and
all the occupied orbitals being degenerate. This excludes, for the 
independent-particle Wannier functions, the use of the mathematical theorems 
onto which the following discussion is based, and also implies that in general 
the functions obtained from the many-body Wannier functions in the limit of no 
interaction differ from the usual non-interacting Wannier functions. 

Also the expression for the many-body density matrix in terms of the many-body 
Wannier functions is slightly more complicated than the corresponding 
expression in the independent-particle case.
\begin{displaymath}
D({\bf r},{\bf r}')
 = \sum_{j,{\bf R},{\bf R}'} n_j({\bf R}-{\bf R}')
   W^\ast_j({\bf r}-{\bf R}) W_j({\bf r}'-{\bf R}') \:, 
\end{displaymath}
where $n_j({\bf R}-{\bf R}')=V/(2\pi)^3\,\int d{\bf k}\,
e^{i{\bf k}({\bf R}-{\bf R}')} n_j({\bf k})$, while for independent 
particles only the term $n_j(0)$ survives.

From a mathematical point of view we can now distinguish two cases. In the 
first case $D_{\bf k}({\bf r}, {\bf r}')$ is an analytic function of ${\bf k}$,
in the second case $D_{\bf k}({\bf r}, {\bf r}')$ has a discontinuity at some
${\bf k}={\bf k_F}$. Then, in the absence of degeneracies, in the first case, 
the occupation band-structure $n_j({\bf k})$ will be an analytic in 
${\bf k}$ \cite{math}, while in the second case it will have a discontinuity 
at ${\bf k_F}$.
We associate the first case with an insulator, while we call systems
with a discontinuity at ${\bf k_F}$ a metal. To justify this classification
we remark that experimentally the sharpness of the Fermi surface in metals 
is strongly suggested by the de Haas-van Alphen effect, cyclotron resonances, 
and Friedel oscillations. Additional support comes from the momentum
distribution $N({\bf p})$. It can be probed by Compton scattering experiments 
\cite{compton}. If $N({\bf p})$ is smooth the system is an insulator, while 
for a metal it shows a discontinuity. For jellium this discontinuity has been 
thoroughly investigated by many-body techniques \cite{daniel}. Being a 
one-particle property, the momentum distribution can be determined from the 
reduced density matrix
\begin{displaymath}
N({\bf p}) = \int d{\bf r} d{\bf r}'\;
             e^{- i {\bf p} ({\bf r}-{\bf r}')} \: D({\bf r},{\bf r}') \:.
\end{displaymath}
Writing the density matrix in terms of the natural orbitals one obtains
\begin{displaymath}
N({\bf p}) =  \sum_{j,\bf G} n_j({\bf p}-{\bf G}) 
             \left| \alpha_{j}^{{\bf p}-{\bf G}}(\bf G) \right|^2 \:, 
\end{displaymath}
where $\alpha_{j}^{{\bf k}}(\bf G)$ are the plane wave expansion coefficients 
of  $U_{j,{\bf k}}({\bf r})$. Obviously $N({\bf p})$ is analytic if 
$n_j({\bf k})$ and $\alpha_{\bf G}^{{\bf k}}$ are analytic, whereas a
discontinuity in $n_j({\bf k})$ manifests itself in a discontinuity in 
$N({\bf p})$. 

We are now in the position of making statements about the decay properties
of the density matrix. 
For an insulator, where the density matrix is analytic in ${\bf k}$, we find 
from (\ref{fdm}) that, since the Fourier transform of an analytic function 
decays exponentially \cite{expd}, the density matrix $D({\bf r},{\bf r}')$ 
decays exponentially fast to zero with increasing distance between ${\bf r}$ 
and ${\bf r}'$. 
For a metal, on the other hand, $D_{\bf k}({\bf r},{\bf r}')$ has
a discontinuity on the Fermi surface and hence the density matrix only decays
algebraically. The decay properties of the density matrix for interacting
systems are thus qualitatively the same as for non-interacting systems. 
We emphasize that this results is valid for zero temperature, while at finite 
$T$ the decay should become exponential also for a metal \cite{finiteT}.

\begin{figure}[ht]
     \begin{center}
      \setlength{\unitlength}{1cm}
      \begin{picture}( 6.,5.8)           
        \put(-1.9,-1.2){\includegraphics{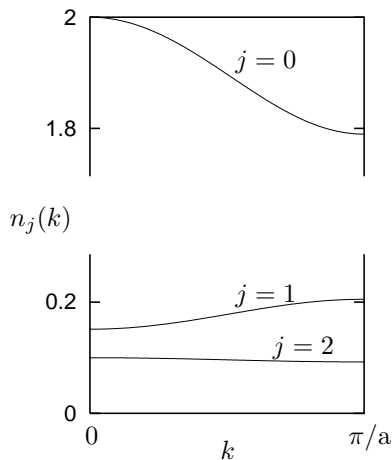}}   
        \put(1.0,0.3){0} 
        \put(2.8,0.1){$k$} 
        \put(0.0,3.2){$n_j(k)$} 
        \put(4.5,0.3){$\pi$/a} 
        \put(3.,5.3){$j=0$} 
        \put(3.,2.2){$j=1$} 
        \put(3.5,1.5){$j=2$} 
       \end{picture}
       \caption{\label{insulator} Qualitative sketch of the 
                occupation band structure $n_j(k)$ for a weakly correlated 
                one-dimensional band insulator with lattice constant $a$.}
      \end{center}
     \end{figure}

We next turn to the properties of the Wannier functions using the same 
mathematical theorems. 
For an insulator, the eigenfunctions $U_{j,{\bf k}}$ are
analytic functions of ${\bf k}$, inheriting their analyticity from 
$D_{\bf k}({\bf r},{\bf r}')$ by  eqn.~(\ref{eveq}). 
Thus by eqn.~(\ref{mbw}) all the many-body
Wannier functions will be exponentially localized for nondegenerate 
occupation bands.
To make more specific predictions about the qualitative shape of the Wannier
functions and to compare to the independent-particle case we will for the moment
specialize our discussion to weakly correlated insulators, i.e.\ correlated
insulators derived from band insulators. Then the occupation numbers of the
(formerly) occupied natural orbitals are close to two (one without spin
degeneracy), while the occupation numbers of the (formerly) virtual natural
orbitals are close to zero. Hence the occupation band structure is fairly
flat. For a one-dimensional band insulator the occupation band structure has 
the qualitative form shown in Fig.~\ref{insulator}. 

We will now put forward some qualitative predictions about the shape of 
the many-body Wannier functions of weakly correlated band insulators. 
As in the independent-particle case such 
a discussion is made difficult by the fact the the Wannier functions 
are not uniquely defined. Multiplying the natural orbitals by 
a ${\bf k}$ dependent phase 
$\Phi_{j,{\bf k}}({\bf r}) \rightarrow 
\Phi_{j,{\bf k}}({\bf r}) \exp(i \theta({\bf k})) $ generates a new set of 
valid natural orbitals leading via eqn.~(\ref{mbw}) to a new set of Wannier 
functions. In the following discussion we will assume that the phase 
$\theta({\bf k})$ is chosen such as to minimize the overall variation 
with respect to ${\bf k}$, i.e. we consider maximally 
localized Wannier functions~\cite{marzari}. 

Eqn.~(\ref{eveq}) shows that the variation of $n_j({\bf k})$ is 
coupled to the one of $\Phi_{j,{\bf k}}({\bf r})$.
We use here the word variation 
in a loose sense, meaning that either the function or its derivatives are large. 
Alternatively, a function has a strong variation if there are regions in which 
the convergence radius of its Taylor expansion is small.
The fact that the occupation band structure is smooth consequently implies 
that the ${\bf k}$ variation of $\Phi_{j,{\bf k}}({\bf r})$ is slow as well. 
The typical length scale over which the variation takes place in ${\bf k}$ 
space is $\pi/a$. The decay constant 
of the many-body Wannier functions is therefore of the order of the interatomic 
distance independent of the index $j$. 

The occupied independent-particle Wannier functions have the same qualitative 
decay behavior. This follows by considering the eigenvalue equation (\ref{bnd}).
Since the occupied bands have in 
general a slow variation over the whole Brillouin zone, the decay constant is 
again of the order of the interatomic distance. This is also what is actually
seen in numerical calculations showing that the occupied natural orbitals
$\Phi_{j,{\bf k}}({\bf r})$ are very similar to the
occupied Kohn-Sham functions $\phi_{j,{\bf k}}({\bf r})~\cite{needs}$.

The virtual independent-particle Wannier functions, however, are more extended 
than the occupied independent-particle Wannier functions. This follows from 
the fact that the higher virtual bands $\epsilon_j({\bf k})$ become free 
particle like and have a strong variation close to the boundaries of the 
Brillouin zone. This trend was also 
demonstrated by explicit calculations~\cite{marzariext}. 

We conclude that the occupied independent-particle and many-body Wannier 
functions of weakly correlated band insulators are very similar. The virtual 
Wannier functions, however, differ in the two contexts. The many-body Wannier 
functions are all localized within the same volume, whereas their 
independent-particle counterparts become more and more extended with increasing
energy.  A similar effect is well known in atoms, where the natural orbitals 
are all localized within the same volume whereas the virtual 
independent-particle orbitals form a Rydberg series with larger and 
larger extent. 

\begin{figure}[ht]
     \begin{center}
      \setlength{\unitlength}{0.85cm}
      \begin{picture}( 6.,7.0)           
        \put(-3.3,-1.2){\includegraphics{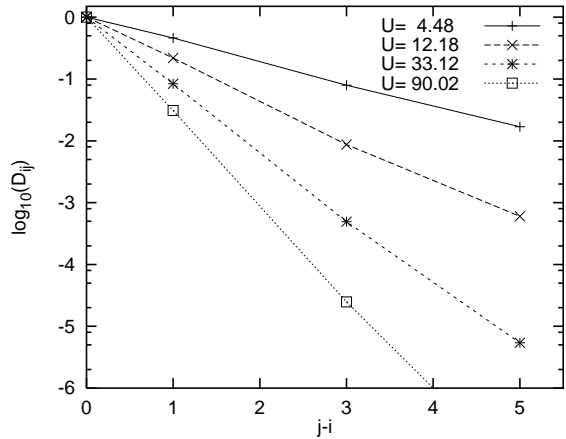}}   
       \end{picture}
    \caption{\label{log12} Exponential decay of the density matrix $D_{i,j}$
             as a function of the inter-site distance $j-i$ (in units of the 
             lattice constant) for a Mott insulator.}
    \end{center}
   \end{figure}

As an example of a strongly correlated insulator we consider a one-dimensional
Hubbard model \cite{hubbard} at half filling. This system is a Mott insulator
\cite{mott} for any finite value of the on-site repulsion $U$ \cite{lieb}.
We have calculated the density matrix for a finite system of 12 sites with
periodic boundary conditions. The results are shown in Fig.~\ref{log12}. 
The exponential decay of the density matrix is clearly visible, being faster
the stronger the correlation, i.e.\ the larger the Hubbard term $U$.
The size of the gap for the systems shown in the figure are given in table 
\ref{gap}.

\begin{figure}[ht]
     \begin{center}
      \setlength{\unitlength}{0.80cm}
      \begin{picture}( 6.,6.4)           
        \put(-2.3,-0.8){\includegraphics{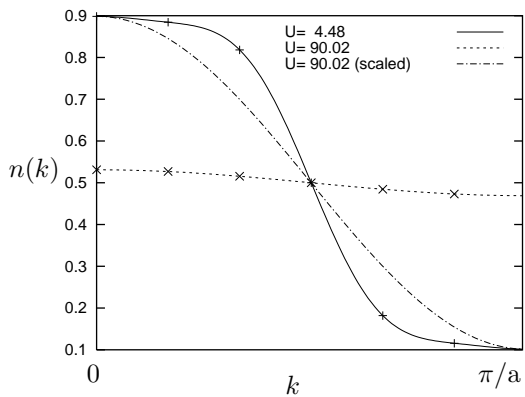}}   
        \put(.0,0.3){0}   
        \put(3.3,0.1){$k$}
        \put(-1.3,3.7){$n(k)$}
        \put(6.5,0.3){$\pi$/a}
       \end{picture}
       \caption{\label{mynk} Occupation band structure of Mott 
        insulators corresponding to two different values of $U$. 
        Clearly the variation of $n(k)$ is stronger for smaller 
        values of $U$. This is true even if the two curves are
        scaled such that they coincide at the end points. For smaller
        correlation $n(k)$ is flat at the edges and decays mainly 
        around $k_F$, over a distance that is roughly half of the 
        width of the cell, while for larger $U$ the decay is nearly
        uniformly over the whole cell. The shorter length scale of
        the first curve manifests itself in a larger decay length
        of the density matrix.}
      \end{center}
     \end{figure}

The occupation band-structure associated with the two extreme $U$-values 
of Fig.~\ref{log12} is shown in Fig.~\ref{mynk}. We notice that, again, a 
strong variation of the occupation band-structure implies a slower decay of 
the density matrix.

\begin{table}[ht]
\begin{center}
\begin{tabular}{cdd}
 $n$ & \multicolumn{1}{c}{$U=\exp{(n/2)}$} & \multicolumn{1}{c}{$E_g$}\\ \hline
 3 &  4.4817 &  2.090 \\
 5 & 12.1825 &  8.840 \\
 7 & 33.1155 & 29.385 \\   
 9 & 90.0171 & 86.142 \\
\end{tabular}
\caption{\label{gap}
 Gap $E_g=E(N-1)+2E(N)-E(N+1)$ for a periodic Hubbard chain of 12 sites
 for various values of $U$. All energies are given in units of the hopping 
 matrix element, i.e.\ the band-width of the non-interacting system is 4.
}
\end{center}
\end{table}

We have shown that the characterization of metals and insulators
by the decay properties of the density matrix can be carried over to 
interacting systems. In addition we have defined many-body Wannier functions 
and discussed their localization properties. We expect that such Wannier 
functions will turn out to be useful in the construction of many-body O(N) 
schemes.
Localization concepts are a basic ingredient for the understanding of
bonding in solids and molecules. Our work extends these concepts that are
well established in non-interacting framework to true many-body systems.

S.G. thanks Nick Trefethen for pointing out references, and T. Arias, 
P. Horsch, O. Gunnarson, G. Stollhoff, J. Hutter, and D. Vanderbilt for 
interesting discussions.

\end{multicols}
\end{document}